\documentclass[12pt]{iopart}

\usepackage{graphicx}
\usepackage{comment}
\usepackage{iopams}
\usepackage{color}

\begin{document}

\title[]{9 GeV Energy Gain in a Beam-Driven Plasma Wakefield Accelerator}

\author{M Litos$^1$, E Adli$^2$, J M Allen$^1$, W An$^3$, C I Clarke$^1$, S Corde$^4$, C E Clayton$^3$, J Frederico$^1$, S J Gessner$^1$, S Z Green$^1$, M J Hogan$^1$, C Joshi$^3$, W. Lu$^5$, K A Marsh$^3$, W B Mori$^3$, M Schmeltz$^1$, N Vafaei-Najafabadi$^3$, V Yakimenko$^1$}

\address{$^1$SLAC National Accelerator Laboratory, Menlo Park, CA, USA}
\address{$^2$University of Oslo, Oslo, Norway}
\address{$^3$University of California Los Angeles, Los Angeles, CA, USA}
\address{$^4$LOA, ENSTA ParisTech, CNRS, Ecole Polytechnique, Université Paris-Saclay, Palaiseau, France}
\address{$^5$ Department of Engineering Physics, Tsinghua University, Beijing, 100084, China}

\ead{litos@slac.stanford.edu}

\begin{abstract}

An electron beam has gained a maximum energy of 9~GeV per particle in a 1.3~m-long electron beam-driven plasma wakefield accelerator. The amount of charge accelerated in the spectral peak was 28.3~pC, and the root-mean-square energy spread was 5.0\%. The mean accelerated charge and energy gain per particle of the 215 shot data set was 115~pC and 5.3~GeV, respectively, corresponding to an acceleration gradient of 4.0~GeV/m at the spectral peak. The mean energy spread of the data set was 5.1\%. These results are consistent with the extrapolation of the previously reported energy gain results using a shorter, 36~cm-long plasma source to within 10\%, evincing a non-evolving wake structure that can propagate distances of over a meter in length. Wake-loading effects were evident in the data through strong dependencies observed between various spectral properties and the amount of accelerated charge.

\end{abstract}


\maketitle

\section{Introduction: Plasma Wakefield Accelerator}

Plasma wakefield acceleration is a method of accelerating charged particles by utilizing the strong electric fields associated with relativistic plasma waves that can be generated in a high-density plasma. The basic scheme involves a high density, ultra-relativstic electron ``drive beam'', which excites an oscillation of the plasma electrons as it passes through the plasma source~\cite{chen1985prl,ruth1985pa,bingham2004ppcf}. In the blowout regime, a region behind the drive beam is made completely devoid of electrons, leaving behind the massive, stationary ions~\cite{rosenzweig1991pra,lu2006pop,lu2006prl}. The size of this ion cavity is on the order of one plasma wavelength in all dimensions. The wake structure travels with the drive beam at a phase velocity approximately equal to the speed of light in a non-evolving, quasi-steady state until: the beam reaches the end of the plasma source; the drive beam is entirely etched away by the head erosion process~\cite{blumenfeld2007nat}; or the drive beam is fully depleted of energy. Within the plasma wake structure are large amplitude electric fields -- wakefields -- along the axis of the beam that can accelerate an appropriately placed second bunch of electrons (the ``trailing beam''). The mechanism can be described as a compact beam energy transformer, increasing the voltage of the trailing beam at the expense of that of the drive beam. Energy is first transferred from the drive beam to the plasma wake, and thence to the accelerated trailing beam, all in a short distance and with a high energy transfer efficiency. The high-gradient, electron-beam driven plasma wakefield accelerator (PWFA) was experimentally demonstrated at SLAC in a series of experiments that initially used a singe electron bunch to both excite and probe the wake~\cite{blumenfeld2007nat,hogan2005prl,muggli2010njp,muggli2004prl}, and then in 2013 by using an electron drive bunch to produce or ``drive'' the wake and a second, distinct electron bunch trailing behind the first bunch to extract energy from the wake~\cite{litos2014nat}. More recently, a variant of this scheme has been shown to work for positrons, as well, where a single positron bunch was shown to excite a plasma wake in which the energy of the particles in the front half of the bunch was efficiently transferred at a high accelerating gradient to positrons in the back of the same bunch creating two spectrally distinct populations of charge~\cite{corde2015nat}. This high gradient, high efficiency acceleration of electrons and positrons is extremely desirable for high energy physics applications, such as a future linear collider~\cite{joshi2003pt,adli2013a,downer2014nat,piot2015nat}. 

The first demonstration of the high-gradient plasma wakefield accelerator (PWFA) using a drive-bunch/trailing-bunch scheme performed at SLAC's Facility for Advanced Accelerator Experimental Tests (FACET)~\cite{hogan2010njp} yielded an average energy gain for the trailing bunch electrons of 1.6~GeV per electron~\cite{litos2014nat}. That experiment used a laser-ionized lithium vapor plasma source~\cite{litos2014nat,green2014ppcf,muggli1999tps} with a density of $5\times10^{16}$~cm$^{-3}$ that was 36~cm long. An accelerating gradient of 4.4~GeV/m was observed in the experiment. The data presented here were collected at FACET in 2014 with a goal of checking the length scaling of the energy gain under the assumption of a non-evolving wake. A non-evolving wake is defined as an accelerating cavity that has a self-similar field structure and a constant phase velocity until the wake is terminated. The experimental setup -- including the plasma density and the drive/trailing bunch parameters -- were nearly identical to the previous round of experiments, except that the lithium vapor plasma source was lengthened to 130~cm and it was ionized by the transverse field of the drive beam~\cite{blumenfeld2007nat,hogan2005prl,muggli2010njp,oconnell2006prstab,ammosov1986spj}, rather than using a laser, as was previously the case. The plasma density was thus $5\times 10^{16}$~cm$^{-3}$, as before, and the properties of the incoming electron beam (charge, spot size, bunch length, and bunch separation) were quite similar in both cases. Therefore, if the wake is non-evolving, then the observed average accelerating gradient (i.e.\ energy gain per unit length) should be the same in both experiments. The expectation would be an increase in the energy gained by the accelerated electron bunch that scales with the ratio of the lengths of the two plasma sources, and the results we observed indeed followed this scaling. We observed an average accelerating gradient of 4.0~GeV/m, which agrees to within 10\% of the previously recorded value of 4.4~GeV/m, where the difference is likely due to the fact that the more recent experiment presented here did not use a pre-ionized plasma source, thus the entirety of the drive beam current could not be utilized for driving the wake. Further, there was likely some degree of beam head erosion over this length, similar to what was observed in previous experiments at SLAC~\cite{blumenfeld2007nat}. The single best shot in our data experienced an average accelerating gradient of 6.9~GeV/m, gaining 9.0~GeV of energy per particle over 1.3~m of plasma.

\section{Methods}

FACET provides a 20.35~GeV electron beam with a two-bunch structure to the plasma source: one bunch to drive the wake, the other bunch trailing behind it to be accelerated by the wake. The methods used to produce and diagnose the two-bunch beam are described in Ref.~\cite{litos2014nat}. The plasma source itself is created from a lithium vapor contained in a heat-pipe oven~\cite{green2014ppcf,muggli1999tps}. From the uniformity of the temperature measurements along the length of the oven we expect that the neutral density of the vapor was constant over the length of the oven at a value of $5\times 10^{16}{\rm cm}^{-3}$, with a density roll-off at the entrance and exit interfaces that had a length of 15~cm. The length of the lithium vapor region for these experiments was 130~cm, as determined by the full-width at half the maximum (flat-top) density. The vapor was ionized by the strong transverse electric fields of the drive bunch~\cite{blumenfeld2007nat,hogan2005prl,muggli2010njp,oconnell2006prstab,ammosov1986spj}, generating a singly-ionized plasma column with an electron density equal to the neutral density of the lithium vapor~\cite{bruhwiler2003pop}. The one second time interval between shots is sufficient to allow the ionized vapor to recombine and reach thermal equilibrium. The thermodynamic properties of the heat pipe oven remain stable over hours, thus the effective plasma length remains consistent throughout the data series, which is recorded over a period of minutes. The bulk of the drive bunch, being located behind the ionization front, excited a non-linear, blowout wake structure that encompassed and accelerated the trailing bunch~\cite{tzoufras2008prl,rosenzweig1988prl}.

The two bunches traveled at ultra-relativistic speeds with no appreciable phase slippage between one another over the full distance of the plasma source, separated by a typical distance of 128$\pm$83~$\mu$m as measured with a radio-frequency transverse deflecting structure that acts as a streak camera by directly streaking the electron beam itself (see online methods section in~\cite{litos2014nat}). The large spread in bunch-to-bunch separation is ultimately attributable to klystron phase jitter in the main linac at the time of the experiment. The method by which the two-bunch electron beam structure is formed at FACET is sensitive to the variation in the beam chirp caused by the klystron phase jitter (see Ref.~\cite{litos2014nat}). Realistic applications would incorporate unique sources for the drive and trailing bunches, and thus the PWFA would not be sensitive to the incoming beam chirp in this way.

The drive and trailing bunches initially contained about 594$\pm$13~pC and 347$\pm$15~pC charge at the entrance to the plasma source, respectively. The rms lengths of the drive and trailing bunches were $\sim$30~$\mu$m and $\sim$50~$\mu$m, respectively, and the projected spot size of the two bunches combined at the plasma entrance was approximately 30~$\mu$m, on average. Due to the large longitudinal extent of the beam, not all of the charge was able to couple into the plasma wake and participate in the beam-plasma interaction. Thus, the final amount of accelerated charge was significantly lower than the input charge of the trailing beam.

\subsection{Imaging Spectrometer}
The final spectrum of the beam was diagnosed with a magnetic imaging spectrometer~\cite{litos2014nat,corde2015nat,adli2015nim} consisting of a quadrupole magnet doublet and a strong dipole magnet that dispersed the beam in the vertical ($y$) plane. The quadrupole magnet strengths were set to image particles of a specific energy from the exit plane of the plasma source to a set of diagnostic screens at the end of the beamline. In order to fully characterize the outgoing beams' spectrum in aggregate, the quadrupoles were adjusted between sub-sets of shots to image different energies from the nominal object plane (i.e.\ the plasma exit location). In the data analyzed here, 100 shots each were taken with the spectrometer set to image energies corresponding to $E_0$, $E_0$+1.5~GeV, $E_0$+3.0~GeV, $E_0$+4.5~GeV, $E_0$+6.0~GeV, and $E_0$+7.5~GeV, where $E_0$ is the initial beam energy of 20.35~GeV. The magnification of the beam in the dimension transverse to the dispersive plane ($x$) was 8.1, while the magnification of the beam in the dispersive plane ($y$) was set to the smaller value of 1.0 in order to optimize the spectral resolution. These constraints required that the imaged energy in the $y$-plane from the plasma exit location to the screens be on the order of 0.5--1~GeV higher than that of the $x$-plane, depending on the imaging condition and the screen location.

The transverse projection of the dispersed beam was recorded at two different screens located at a distance $L=9.57$~m and $L=10.9$~m downstream of the dipole magnet, corresponding to a dispersion value $D_0$ for electrons at the energy $E_0$ of 54.8~mm and 59.5~mm, respectively. The dispersion $D_0=\Delta \theta_0 \cdot L$, and $\Delta \theta_0=5.73$~mrad is the deflection angle for electrons at the energy $E_0$. The first screen was a phosphorescent sheet of Lanex material affixed to the downstream side of a 1~mm thick converter foil made of tungsten, viewed by a 16-bit scientific CMOS camera. The second screen was comprised of a silicon wafer reflecting Cherenkov light emitted by the passage of the electron beam through a 10~cm air gap immediately preceded by another silicon wafer used to block the Cherenkov light emitted further upstream~\cite{adli2015nim}. The Cherenkov light was collected by a 16-bit scientific CMOS camera. The Cherenkov screen was roughly ten times less sensitive than the Lanex screen, but it covered a wider field of view and was insensitive to synchrotron, betatron, and Bremsstrahlung radiation backgrounds, which can be problematic under certain experimental conditions (though these effects were not a significant source of background in this study). The Cherenkov screen was used as the primary diagnostic for this study, while the Lanex screen was used as a secondary check for the major spectral features in the beam. The spectral features in the data were analyzed in two different ways for every shot: one method focused on the spectral charge density peak, and the other focused on the centroid energy value of all the accelerated charge in a given shot.

\subsection{Spectral Analysis: Charge Density Peak Energy}\label{sec:spectral peak}
The spectral charge density (charge within a given energy bandwidth) peak (see Figure \ref{fig:fig1}a) was defined as the energy at which the projected spectral charge density was the highest for any point in the region above 21.5~GeV, which roughly delineates the upper reach of the slightly accelerated tail of the drive beam. The energy spread for the spectral charge density peak was calculated as the root-mean-square value for the distribution of charge above the spectral peak energy (also shown in Figure \ref{fig:fig1}a). The upper limit of the domain of the root-mean-square calculation was the point below which 95\% of all observed charge above the spectral peak was located. To estimate the accelerated charge at the spectral peak, we summed all the charge located within two times the rms energy spread of the spectral peak. The contrast ratio of the peak was calculated as the ratio between the peak charge density and the charge density of the trough between $E_0$ and the spectral peak. In order to be considered for this analysis, the contrast ratio was required to be larger than 1.5 for a given shot. This removed 385 shots from the complete 600 shot data set. The spectral peak-based approach to the analysis is important because it places more emphasis on what would be considered the more ``usable'' portion of the beam by mostly excluding the lower energy tail of the accelerated charge. On average, the accelerated charge about the peak accounts for roughly 84\% of that considered when using the centroid energy calculation, described below. As an example, the spectral peak and the rms energy spread are shown for a shot with 9 GeV peak energy gain in Figure~\ref{fig:fig1}a with solid and dashed black lines, respectively.

\subsection{Spectral Analysis: Centroid Energy}
The centroid (median) energy value was calculated using the entire range of energy extending above the spectral charge density trough. The energy spread about the centroid was calculated as the root-mean-square value over a domain that extended outward from the centroid until 95\% of all charge observed above the spectral charge density trough was included, keeping the lower limit bounded at the spectral trough energy. The total accelerated charge about the centroid energy was calculated as the sum of charge located within two times the rms energy spread of the centroid energy, again with a lower bound at the spectral charge density trough. Because there was often a low-charge density tail that was accelerated to a lesser extent along with the mass of charge at the spectral peak, the centroid energy value tends to be slightly lower than the spectral charge density peak energy value, though the two track one another closely, giving energy gain values that agree to within 15\% on average (see Figure \ref{fig:fig2}). The centroid energy analysis is valuable because it allows us to consider the entirety of the accelerated charge, which is needed to extract causal relationships in the final spectrum related to wake loading effects, as discussed in the Section 3.

\section{Results: Spectrum of Accelerated Beam}

After rejecting the 385 shots with low contrast, we analyzed the remaining 215 of the recorded 600 sequential shots that had a distinct spectral charge density peak in the final energy projection of the beam as discussed in section \ref{sec:spectral peak}. The remainder of the shots had either a continuous energy distribution, or experienced a weak plasma interaction with little to no effect on the beam's final energy. The 215 peaked shots were distributed evenly throughout the data set, where the variance of the beam-plasma interaction was dependent on the variation of the incoming electron beam parameters. Specifically, the shot-by-shot jitter of the energy content of the incoming beam affected the relative transverse alignment and longitudinal separation of the drive and witness bunches at the plasma entrance due to dispersive effects introduced by the magnetic beam delivery system. The relative transverse alignment jitter was directly observed on a shot-by-shot basis for the data collected by way of an optical transition radiation signal produced by the passage of the electron beam through a 1~$\mu$m thick titanium foil located a few meters upstream of the plasma entrance. Most significantly, the standard deviation of the bunch-to-bunch longitudinal separation was on the order of half the plasma wavelength, as reported in Section 2.

\begin{figure}[ht!]
  \centering
 \includegraphics[width=0.95\textwidth]{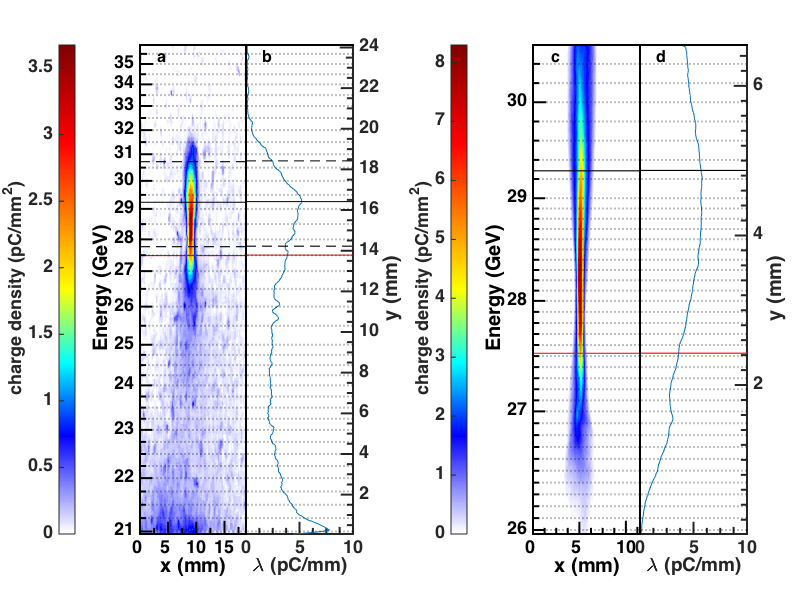}
  \caption{(a) and (c) show the energetically dispersed transverse charge density profile of the highest peak energy shot from the data set as observed on the wide-field of view (FOV) Cherenkov screen and the Lanex screen, respectively. The left-axis displays the energy calibration of the screen, and the right and bottom axes display the physical size of the beam on the screen. The color axis corresponds to the charge density in units of pC/mm$^2$, represented on a linear scale. The horizontal lines represent centroid energy (red), the peak energy (solid black), and the values corresponding to the rms energy spread about the peak energy (dashed black). All of these values were calculated for the Cherenkov screen shown in (a). (b) and (d) show the horizontally integrated spectral charge density profiles from (a) and (c), respectively.}
  \label{fig:fig1}
\end{figure}

\subsection{Highest Energy Gain}
Figure~\ref{fig:fig1}a and b show the energetically dispersed profile of the single shot from the data which had the highest spectral peak energy, 29.3~GeV, corresponding to an energy gain per unit length of 6.9~GeV/m. The amount of charge accelerated in the peak of this shot was 28.3$\pm$2~pC, and the energy spread about the peak was 5.0\%. Figure~\ref{fig:fig1}c and d show the same shot as recorded on the Lanex screen, which gives a spectral peak value of 29.2~GeV, agreeing with the wide FOV Cherenkov screen to better than 3\%. The centroid energy, energy spread, and accelerated charge values for this shot were 27.5~GeV, 8.0\%, and 43$\pm$5~pC, respectively. 

\begin{figure}[ht!]
  \centering
 \includegraphics[width=0.95\textwidth]{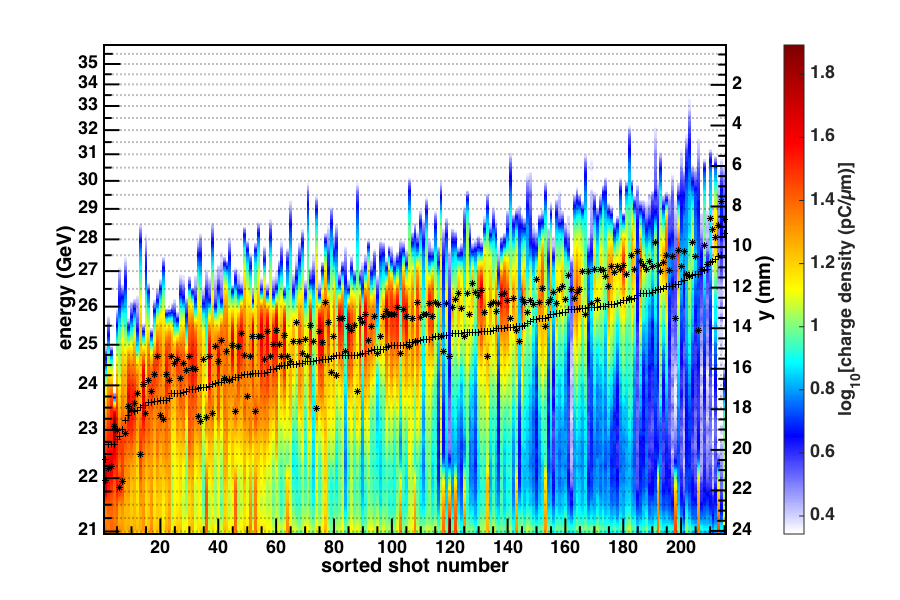}
  \caption{A plot of the charge-integrated spectral projection of all 215 analyzed shots, sorted in order of ascending centroid energy value. The color axis represents the spectral charge density on a logarithmic scale. The star-shaped markers indicate the peak-energy; the cross-shaped markers indicate the centroid energy.}
  \label{fig:fig2}
\end{figure}

\subsection{Statistical Spectral Analysis}
The projected spectral charge density of all 215 analyzed shots are shown in Figure~\ref{fig:fig2}, where the charge density is represented on a log-scale by the color axis. The shots are sorted from left to right in order of ascending centroid energy, which is depicted by the cross-shaped markers, while the spectral peak of each shot is indicated by the star-shaped markers. The distributions of energy gain, energy spread, and accelerated charge values as calculated for the centroid energy and peak energy all roughly fall into normal distributions. A statistical summary of these quantities is presented in Table~\ref{tab:tab1}.

\begin{table}
\caption{\label{tab:tab1} Statistical analysis of accelerated beam spectra, including the standard deviation (s.d.) of each measured quantity. Values are given for calculation techniques using both the centroid energy and spectral peak energy.}
\footnotesize\rm
\begin{tabular*}{\textwidth}{@{}l*{15}{@{\extracolsep{0pt plus12pt}}l}}
\br
Measured Quantity & Centroid Energy & Spectral Peak Energy \\
\mr
Mean Energy Gain & 4.7~GeV (1.1~GeV s.d.) & 5.3~GeV (1.4~GeV s.d.) \\
Mean RMS Energy Spread & 5.9\% (1.3\% s.d.) & 5.1\% (2.3\% s.d.) \\
Mean Accelerated Charge & 140~pC (55~pC s.d.) & 120~pC (47~pC s.d.) \\
\br
\end{tabular*}
\end{table}

The mean accelerating gradient for the spectral peak energy was 4.0~GeV/m, which is within 10\% of the previously recorded value of 4.4~GeV/m in the previous two-bunch PWFA experiments at SLAC in Ref.~\cite{litos2014nat}, indicating that the plasma wake is non-evolving over meter-scale distances. The mean total energy transferred from the drive beam via the wake to the accelerated charge in the the experimental results presented here was 0.6~J, which is greater than the energy transfer in the previous experiment by a factor of about five. This difference can be accounted for by the ratio of the length of the two plasma sources (3.6) and the ratio of accelerated charge (1.6), the combination of which would lead to a rough estimate of an improvement in energy transfer of about a factor of 5.8, which is within 16\% of the experimental value reported here.

\begin{figure}[ht!]
  \centering
 \includegraphics[width=0.95\textwidth]{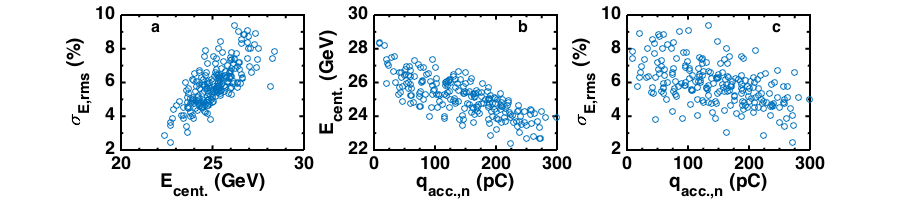}
  \caption{(a) Scatter plot showing the correlation between the energy spread and the centroid energy. (b) Scatter plot showing the anti-correlation between the centroid energy $E_{cent.}$ and the weighted accelerated charge $q_{acc.,n} = q_{acc.}\cdot <E_{max.}>/E_{max}$, where $q_{acc.}$ is the raw accelerated charge value for a given shot, and the weighting factor $<E_{max.}>/E_{max}$ is the mean maximum energy observed divided by the maximum energy for the particular shot. (c) Scatter plot showing the anti-correlation between the rms energy spread about the centroid energy $\sigma_{E,rms}$ and the weighted accelerated charge. In both (b) and (c) the maximum energy stands in as a proxy for the strength of the wakefield.} 
  \label{fig:fig3}
\end{figure}

\subsection{Wake Loading Effects}
The variation of the incoming electron beam allowed us to sample a range of final outcomes. In particular, as the amount of accelerated charge was varied, we could observe a closely correlated change in the final spectrum. As the charge increased, the energy spread and the centroid energy decreased as shown in Figure~\ref{fig:fig3}b and c. The physics of wake loading implies this to be a causal relationship. For a non-linear plasma wake that is loaded with an ideal beam that perfectly flattens the wakefield, the product of the accelerated charge and the energy gain is conserved~\cite{tzoufras2008prl}. That we have a large energy spread in our final spectrum (see Figure~\ref{fig:fig3}a and c) is an indicator that we are not perfectly flattening the plasma wake, though we still expect that the energy available for acceleration inside the wake remains the same from shot-to-shot, on average, and thus by loading it down with more charge we expect to see the mean energy gained per particle in a given shot to decrease in a roughly linear fashion. 

In Figure~\ref{fig:fig3} we have plotted three strong correlations that give evidence of wake-loading effects in the data. Figure~\ref{fig:fig3}a shows the increase of energy spread with the increase of centroid energy gain. This observation on its own gives a strong suggestion of wake loading, but there is no causation implied by the variables plotted alone. To see evidence of the cause for the correlation in Figure~\ref{fig:fig3}a, we must turn to Figure~\ref{fig:fig3}b and c. In both of these plots, the causal variable is the amount of weighted accelerated charge, $q_{acc.,n} = q_{acc.}\cdot <E_{max.}>/E_{max}$. The accelerated charge tells us how much charge was sitting in the accelerating phase of the wake, while the maximum energy reach of a given shot $E_{max}$ serves as a proxy for the strength of the wake. This rests on the assumption that the high energy tail consistently samples the highest field in the wake for every shot, and thus by measuring the highest energy observed in the tail, we can get an indirect handle on the strength of the wake in which it was accelerated for a particular shot. We use the maximum energy as a weighting factor for the accelerated charge in each shot to give a more uniform comparison of the wake loading effects across all the analyzed data.

The wake loading theory tells us that when the wake is well-flattened, the product of the accelerated charge and the energy gain of that charge is held constant~\cite{tzoufras2008prl}. Further, simple conservation of energy tells us that if we add more and more particles to the accelerating phase of the wake then the average amount of energy that can be gained by each must get smaller and smaller. What we observe in Figure~\ref{fig:fig3}b shows exactly this. As the normalized accelerated charge is increased, the centroid energy is reduced, following a fairly linear trend.

Another consequence of the wake loading dynamics is that the energy spread is decreased as the wake is loaded more strongly. This is a behavior that can arise from under-loading the wake, which occurs when there is not sufficient charge to fully flatten the wakefield. When the charge is reduced, the wakefield becomes less flat and each longitudinal slice of the accelerated beam experiences a field strength that is different from the slice next to it.

Put together, the consequences of the wake-loading dependencies shown in Figure~\ref{fig:fig3}b and c explain the observed correlation between the energy spread and the centroid energy gain shown in Figure~\ref{fig:fig3}a. These plots describe a picture where the plasma wake is more loaded down and better flattened when more charge is added to the accelerating phase of the wake, suggesting that the experiment was operated in a somewhat under-loaded regime. The fact that the minimum observed energy spread about the centroid energy was 2.43\% (0.92\% about the peak energy) already tells us that this was the case, as a perfectly loaded wake would lead to no increase in the energy spread beyond the intrinsic energy spread of the witness prior to acceleration in the PWFA.

\section{Summary and Conclusions}
9 GeV energy gain per particle was observed in a 1.3~m long plasma wakefield accelerator operating at a plasma density of $5\times 10^{16}$~cm$^{-3}$, giving an accelerating gradient of 6.9~GeV/m. The statistical mean peak energy gain of 215 analyzed shots was 5.3~GeV, corresponding to a gradient of 4.0~GeV/m. This accelerating gradient agrees to within 10\% of that previously reported in Ref.~\cite{litos2014nat}, which had nearly identical experimental parameters save the length of the plasma source. This consistency is a demonstration that the plasma wake is non-evolving throughout the beam-plasma interaction and can remain in this state for distances greater than one meter. The statistical mean energy spread about the spectral peak was relatively narrow at 5.1\%, strongly suggesting the physics of wake-loading at work. Variation in the amount of accelerated charge and in the plasma wakefield strength (owing to variation in the drive bunch compression and focusing) led to highly correlated changes in the final spectrum. Increased charge in the accelerating phase of the wake gave rise to stronger loading and better flattening of the wake, which reduced the energy gain and energy spread of the accelerated beam. 

\section*{Acknowledgments}
Work at SLAC was supported by DOE contract number DE-AC02-76SF00515. Work at UCLA was supported by DOE grant number DE-SC0010064 and the NSF grant number PHY-1415386.


\end{document}